\documentclass[preprintnumbers,amsmath,amssymb]{revtex4}
\usepackage{amssymb}
\usepackage{amsmath}
\usepackage{epsfig}
\usepackage{graphicx}

\newcommand{\be}{\begin{equation}}
\newcommand{\ee}{\end{equation}}
\newcommand{\bea}{\begin{eqnarray}}
\newcommand{\eea}{\end{eqnarray}}
\numberwithin{equation}{section}
\begin{document}
%\begin{frontmatter}
\title{A note on the path integral representation for Majorana fermions}
\author{Andr\'es Greco}
\affiliation{Facultad de Ciencias Exactas, Ingenier\'{\i}a y Agrimensura and 
Instituto de F\'{\i}sica Rosario
(UNR-CONICET). 
Bv. 27 de Febrero 210 Bis, 2000 Rosario, Argentina}

\begin{abstract}
Majorana fermions are currently of huge interest in the context of 
nanoscience and condensed matter physics. Different to usual fermions, Majorana fermions
have the property that the particle is its own anti-particle thus, they must be described by real fields. 
Mathematically, this property makes nontrivial the quantization of the problem due,
for instance, to the absence of a Wick-like theorem. 
In view of the present interest on the subject, it is important to develop different 
theoretical approaches in order to study problems where Majorana fermions are involved.
In this note we show that Majorana fermions can be studied in the context of field theories for 
constrained systems. 
Using the Faddeev-Jackiw formalism for quantum field theories with constraints, we 
derived the path integral representation for Majorana fermions.  
In order to show the validity of the path integral we apply it to an exactly solvable problem. 
This application also shows that it is rather simple to perform 
systematic calculations on the basis of the present framework.   
\end{abstract}

\maketitle

\vspace*{0.5cm}

%\end{frontmatter}
\section{Introduction}
%

%\bea
%\label{Solution1}
%{\vec \nabla}_{\cal T} \times {\vec {\cal A}}^{\cal T}  =  - \vec {\cal T} \; ,
%\eea
%We notice furthermore that, from the last two equations in (\ref{Solution1}) 

In 1937 Ettore Majorana discovered the real solution of the Dirac 
equation\cite{majorana37}, and he also speculated that his solution 
might describe the physics of the neutrinos. 
Since that time, Majorana fermions are a matter 
of intense studies in particle physics\cite{aste09}.  
However, in the last years great interest on them occurred in condensed matter 
physics\cite{elliott15,wilczek09}. 
For instance, the pioneer works from Kitaev \cite{kitaev03} have pushed the 
topic towards topological phases in spin systems (see \cite{karnaukhov12} and references their in). 
The recent observation of Majorana fermions in 
ferromagnetic chains on a superconductor is also very remarkable\cite{nadj14}.
In this context it is worth to note that in the period 2014-2015
about 500 papers containing the word Majorana in 
the abstract were reported only in the condensed-matter-section of arXiv. 

Majorana fermions have the property that 
the particle is its own anti-particle. This property leads to 
commutation  rules which differ considerably from the commutation rules for usual 
fermions. As it is well known, the commutation rules for fermions are 
a key point for the validity of the Wick theorem, and the consequent theoretical quantum treatment 
of many-body systems. 
Thus, the canonical quantization of Majorana fermions is not trivial.  
An alternative theoretical framework for quantization is via the 
path integral representation. Previous work \cite{nilsson13} focused on this problem
adapting the coherent state formalism \cite{perelomov86}
to Majorana fermions. Since there are not coherent
states for Majorana fermions the derivation can be done
using pairs of conventional fermions.
Thus, the path integral formulation is not straightforward
(see also \cite{shankar11}). The fact that Majorana
fermions is a timely subject it is important to develop
a new alternative method for such derivation

In this note we show that a  theory containing  Majorana fermions can be considered as a 
constrained system. The treatment of constrained systems was initiated 
by Dirac \cite{Dirac64} and continued by Faddeev and Jackiw (FJ) \cite{Faddeev88,Govaerts90}. 
This approach, which is well known in field theory, allows to write the corresponding path 
integral for the model after obtaining the classical effective action consistent with 
the corresponding quantum theory. Thus, the path integral allows to 
study the quantum problem and for instance read the Feynman rules from the effective action\cite{hooft73}. 
The FJ method was used also in solid state physics  
for the Heisenberg model\cite{foussats99a}, $t-J$ model \cite{foussats99b}, and for 
bond-fields in spin systems\cite{foussats11}. 

The paper is organized as follows. In section II we present the Faddeev-Jackiw method for 
Majorana fermions and derive the corresponding path integral representation. In section III we show the validity 
of  the path integral by applying it to a simple and exactly solvable problem. 
In addition, this example shows how to work systematically 
with this formulation. Discussion and conclusion are given in section IV. 
\section{Path integral formulation for Majorana fermions}
\subsection{Quantum algebra for Majorana fermions: difficulties about the 
canonical quantization}

Usual fermions are described by the well known creation and destruction operators $\hat{f}_a^\dag$ and $\hat{f}_a$, respectively,
that satisfy the following anticommutation rules.

\be
\{\hat{f}_a,\hat{f}_b^\dag\}=\hat{f}_a \hat{f}_b^\dag+ \hat{f}_b^\dag \hat{f}_a = \delta_{ab}.
\ee

Mathematically, the property that for Majorana fermions  
the particle is its own anti-particle is expressed 
as $\hat{f}_a^\dag=\hat{f}_a$. Thus, the commutation rules for 
Majorara fermions become 

\be
\label{commM}
\{\hat{f}_a,\hat{f}_b^\dag\}=\{\hat{f}_a^\dag,\hat{f}_b\}=\{\hat{f}_a^\dag,\hat{f}_b^\dag\}=
\{\hat{f}_a,\hat{f}_b\}=\delta_{ab},
\ee

\noindent leading, for instance, to the unusual property $\hat{f}_a\hat{f}_a=1/2$.

The commutation rules (\ref{commM}) makes the quantum problem complicated.
At the operatorial or canonical level 
one way is to work in the Hilbert space generated by $\hat{f}_a^\dag$ and $\hat{f}_a$, and 
the constraint between operators $\hat{\Omega}=\hat{f}_a^\dag-\hat{f}_a=0$ should be applied rigorously. 
At this point it is worth to mention that the commutation rules for Majorana fermions (\ref{commM}) make 
difficult the formulation of a Wick theorem and thus, the quantum many-body treatment is not trivial. 
Instead of using the canonical quantization, in the next subsection we introduce the path 
integral representation for Majorana fermions.  

\subsection{Faddeev-Jackiw theory  and path integral for Majorana fermions\label{PI}}

In order to formulate a path integral representation for quantum systems, 
the coherent-state approach \cite{perelomov86} is 
often used. Instead of that, here we use the formalism introduced by 
FJ \cite{Faddeev88,Govaerts90,Barcelo92} for constrained system, that provides a 
way for obtaining a classical theory  
consistent with the algebra of the quantum problem. 
In this approach, no formal distinction is made between different forms of constraints 
like in Dirac's theory \cite{Dirac64}, where primary and secondary, first class and 
second class constraints appear. In the FJ method \cite{Faddeev88,Govaerts90,Barcelo92} 
the constraints can be incorporated iteratively until the basic brackets 
for the fields can be determined.
Once the classical field-theory is obtained, quantization can proceed 
via a path integral or in a canonical way. It is worth to remember that a path integral 
formulation is written in terms of classical variables, Grassmann or complex regular numbers 
for fermionic or bosonic fields\cite{berezin}, respectively.
Since the FJ-formalism is not widely used, we will briefly discuss 
it. 

Following the Faddeev-Jackiw theory\cite{Faddeev88} and its extension to
fermionic degrees of freedom\cite{Govaerts90}, we 
defined the following effective first-order classical Lagrangian

\be
\label{LFJ}
L=\sum_A \dot{z}_A K^A(z_A) -H(z_A),
\ee

\noindent where $z_A$ are fermionic or bosonic fields, 
$H(z_A)$ is the Hamiltonian, and the coefficients $K^A(z_A)$ are functions of the 
fiels $z_A$ (see below). 

It is possible to prove\cite{Faddeev88} that if the matrix

\be
\label{MAB}
M_{AB}=\frac{\partial K^B}{\partial z_A} - (-1)^{\epsilon_A \epsilon_B} \frac{\partial K^A}{\partial z_B}
\ee

\noindent is not singular, the theory is unconstrained. In (\ref{MAB}) 
$\epsilon_A=1,0$ if $z_A$ is a fermionic or bosonic field, respectively. 
In this case the basic bracket  (or generalized bracket) of the FJ theory, 
defined as
 
\be
\label{FJbra}
\{z_A,z_B\}_{FJ}=(-1)^{\epsilon_A} M^{-1}_{AB},
\ee

\noindent agrees with the Poisson bracket. $M^{-1}_{AB}$ is the element $(AB)$ of the inverse 
of the matrix $M_{AB}$.

If the theory contains a constraint $\Omega(z_A)=0$, this can be included 
using a Lagrange multiplier $\xi$, 

\be
\label{LFJ1}
L=\sum_A \dot{z}_A K^A(z_A) -H(z_A)+\xi \Omega(z_A)
\ee

In this case the matrix $M_{AB}$ is singular and its zero mode encodes information
of the constraint. 

Following \cite{Faddeev88}, the variable $\xi$ disappears  after imposing the constraint, leading to the 
first-iterated Lagrangian defined as

\be
\label{LFJ1}
L=\sum_A \dot{z}_A K^A(z_A)+\dot{\lambda} \Omega(z_A)-H(z_A)
\ee

\noindent In \ref{LFJ1}  the space was enlarge as  $z_A\longrightarrow(z_a,\lambda)$, and
$K^A(z_A)\longrightarrow(K^A(z_A),\Omega(z_A))$.

Now, the new matrix $M_{AB}$ is nonsingular 
and the FJ generalized bracket (\ref{FJbra}) agree with the Dirac bracket \cite{Dirac64} 
of the constrained theory.

As usual, to pass from the classical theory to the quantum theory we have to associate each 
classical variable with the corresponding operator, and impose the 
commutation between them. In the FJ and Dirac theories the prescription  is 

\be
\label{comm}
[\hat{z}_A,\hat{z}_B]_\pm=-{\rm i} \{z_A,z_B\}_{FJ}
\ee

\noindent where $+$ ($-$) is for an anticommutator (commutator).  

Note that we use the hat symbol for quantum operators, while the variables without 
that symbol are classical variables.

It is important to note that (\ref{FJbra})-(\ref{comm}) may be used to obtain the commutation rules of the 
quantum theory when the coefficients $K^A(z_A)$ are known. On the other hand, if the commutation rules are known, as in the case of Majorana fermions, 
(\ref{FJbra})-(\ref{comm}) can be used for deriving 
the explicit 
expression for the coefficients $K^A(z_A)$ introduced in (\ref{LFJ}).

Next, we will review the above ideas applying the FJ formalism to 
our problem of interest. Following (\ref{LFJ1}) we propose the first-iterated 
Lagrangian for the Majorana fermions

\be
\label{LM}
L=\sum_a [\dot{f}_a^\dag A^{f^\dag}_a(f,f^\dag)+\dot{f}_aA^f_a(f^\dag,f)+\dot{\lambda}_a \Omega_a]-H(f^\dag,f)
\ee

\noindent where $f_a$ and $f^\dag_a$ are complex fermionic Grassmann variables, and $\lambda$ is a 
Grassmann Lagrangian multiplier associated with the constraint $\Omega_a=f^\dag_a-f_a=0$. 
In (\ref{LM}) $H$ is a proper model Hamiltonian.

Note that at the classical level $f_a$ and $f^\dag_a$ are considered as Grassmann fields, and for instance 
$f_af_a=0$ due to the 
properties of the Grassman algebra\cite{berezin}, which should not be confused with an operatorial 
rule. 

Identifying the set of variables $(f^\dag_a, f_a,\lambda_a)$, 
and the coefficients $(A^{f^\dag}_a(f,f^\dag),A^f_a(f^\dag,f),\Omega_a)$, the 
explicit expression for the unknown coefficients $A^{f^\dag}_a(f,f^\dag)$ and $A^f_a(f^\dag,f)$
in \ref{LM} can be determined in such a way that the matrix $M_{AB}$ leads to the commutation rules (\ref{commM}) for Majorana fermions.

After some algebra we obtain

\be
A^{f^\dag}_a(f,f^\dag)=\frac{{\rm i}}{4} f_a, 
\ee

\noindent and 

\be
A^{f}_a(f,f^\dag)=\frac{{\rm i}}{4} f^\dag_a, 
\ee

Thus, the explicit expression for $L$ is  

\be
\label{Lef}
L=\sum_a [\frac{{\rm i}}{4} \dot{f}^\dag_a f_a + \frac{{\rm i}}{4} \dot{f}_a f^\dag_a + \dot{\lambda}_a \Omega_a] -H(f^\dag,f)
\ee

It is useful to show that the commutation rules for Majorana fermions are recovered.
Using (\ref{MAB}) $M_{AB}$ can be easily calculated, and its inverse is

%\begin{widetext}
\begin{equation} 
M^{-1}_{AB}= 
\left(
\begin{array}{llllll}
\;\;{\rm i}&\;\;{\rm i}&-\frac{1}{2} \\
\;\;{\rm i}&\;\;{\rm i}&\;\;\frac{1}{2} \\
-\frac{1}{2}&\;\;\frac{1}{2}&\;\;-\frac{{\rm i}}{4}
\end{array}
\right)\delta_{ab} \; ,
\end{equation} 
%\end{widetext}

Thus, the FJ brackets between  $f_a$ and $f^\dag_b$ are

\be
\{f_a,f^\dag_b\}_{FJ}=\{f^\dag_a,f_b\}_{FJ}=\{f^\dag_a,f^\dag_b\}_{FJ}=\{f_a,f_b\}_{FJ}={\rm i}\delta_{ab},
\ee

Therefore, using (\ref{comm}) the commutation rules for 
Majorana fermions are obtained.

Once the classical effective Lagrangian (\ref{Lef})  that leads to the proper 
quantum algebra (\ref{commM}) is obtained, the next step is to write 
the path integral. Following  \cite{Gitman90,senjanovic}

\be
\label{PI1}
\int {\cal D}f {\cal D}f^\dag (det M_{AB})^{1/2} \delta(\Omega) e^{{\rm i} \int dt L'} 
\ee

\noindent where $L'=L\mid_{\Omega=0}$, i.e., 

\be
\label{Lef1}
L'=\sum_a [\frac{{\rm i}}{4} \dot{f}^\dag_a f_a + \frac{{\rm i}}{4} \dot{f}_a f^\dag_a] -H(f^\dag,f),
\ee

\noindent and  $det M_{AB}$ is the determinant of $M_{AB}$. 

In contrast to other theories\cite{foussats99b,foussats11} $det M_{AB}$ does not depend on the fields
and can be absorbed into the measure of the path integral.
The integration on $f^\dag_a$ can be done using the $\delta(\Omega)$. Thus, (\ref{PI1}) is

\be
\label{PI2}
\int {\cal D}f e^{{\rm i} \int dt L_{\rm eff}}
\ee

\noindent where 

\be
\label{Lef2}
L_{\rm eff}=\sum_a \frac{{\rm i}}{2} \dot{f}_a\,{f}_a -H(f)
\ee

%The prefactor $1/4$ in the kinetic term of (\ref{Lef}) is not trivial nd deseved a short dcussion.
%It is well know that the effective Lagrangian that enters in the path 
%integral for usual fermions $f$ is 

%\be
%L=\frac{i}{2} \dot{f}^\dag f+\frac{i}{2} \dot{f} f^\dag -H(f^\dag,f)
%\ee

%Using the Faddeev-Jackiw theory for this case we have a $2\times2$ matrix $M_{AB}$ 
%whose inverse is 

%\begin{widetext}
%\begin{equation} 
%M^{-1}_{AB}= 
%\left(
%\begin{array}{llllll}
%0&i\\
%i&0 
%\end{array}
%\right) \; ,
%\end{equation}
%\end{widetext}

%\noindent which means that 
%\be
%\{f,f^\dag\}=\{f^\dag,f\}=1
%\ee

%\noindent and 
%\be
%\{f,f\}=\{f^\dag,f^\dag\}=0
%\ee

At this point we make two formal remarks: a) The kinetic action for the case of Majorana fermions contains a factor 
$1/4$ (\ref{Lef}) instead of $1/2$ as for usual fermions\cite{auerbach94}. One may intuitively think that 
the effective Lagrangian for Majorana fermions is that of the usual fermions where the 
kinetic action contains a factor $1/2$, plus the constraint $\Omega=0$ 
for imposing the reality of the fields. However, it is possible to show that this proposal 
does not lead to the correct commutation rules (\ref{commM}). 
b) The kinetic term of (\ref{Lef2}) is 
$\frac{{\rm i}}{2} \dot{f} f$. For the case of a real boson $b$, a kinetic term of the form 
$\dot{b} b$ can not exist, because it can be expressed as a total derivative respect to time,  
$ \dot{b} b=\frac{1}{2} \frac{\partial{(b b)}}{\partial{t}}$. Thus, $\dot{b} b$ can be neglected in the effective action. 
However, for Grassmann variables $\frac{\partial{(f f)}}{{\partial{t}}}=\dot{f} f-\dot{f}f=0$, which means that $\dot{f} f$ 
can not be written as a total derivative with respect to time and can not be neglected.

\section{The Majorana propagator and an example of application}

\subsection{An exactly solvable problem}

In this section we apply the path integral representation obtained in 
the last section to a simple and exactly solvable problem. There are two 
aims for this calculation: a) to show that the path integral is well defined and at 
the same time b) to show how systematic calculations can be performed on this framework.  

As in \cite{nilsson13} we propose the following 
Hamiltonian 

\be
\label{Hsimp}
\hat{H}=\epsilon (\hat{c}^\dag \hat{c}-1/2)
\ee

\noindent where $\hat{c}^\dag$ ($\hat{c}$) creates (annihilates) a fermion 
in an orbital of energy $\epsilon$. The energy was trivially shifted by
$-\epsilon/2$ for convenience. Using the standard many-body theory it is straightforward to show that 
$\langle H\rangle =-\frac{\epsilon}{2}$tanh($\frac{\epsilon}{2k_B T}$), where $T$ is the temperature.

It is easy to see that using the commutation rules for 
Majorana fermions,
$\hat{c}$ can be written as $\hat{c}=(\hat{\chi}_1-{\rm i}\hat{\chi}_2)/\sqrt{2}$,
where $\hat{\chi}_1$ and $\hat{\chi}_2$ are two Majorana fermions. 

In terms of $\hat{\chi}_1$ and $\hat{\chi}_2$ 
the Hamiltonian is:

\be 
\label{HMaj}
\hat{H}=-{\rm i} \epsilon \hat{\chi}_1 \hat{\chi}_2
\ee

Following the results of the previous section we first remove 
the hat symbol and introduce the classical Hamiltonian 

\be 
{H}=-{\rm i} \epsilon {\chi}_1 {\chi}_2
\ee

\noindent where now $\chi_1$ and $\chi_2$ are two Grassmann variables.

Second, following (\ref{PI2}), the path integral for this problem reads

\be
\int {\cal D} \chi_1 {\cal D} \chi_2
e^{{\rm i} \int dt L_{\rm eff}} 
\ee

\noindent where the classical effective Lagrangian is 

\be
L_{\rm eff}=\frac{i}{2} (\dot{\chi}_1\,{\chi_1}  + \dot{\chi}_2\,{\chi_2} )+{\rm i} \epsilon \chi_1 \chi_2
\ee

Going to the Euclidean time $\tau={\rm i}t$,
the path integral partition function, which allows the many-body calculation at 
finite temperature\cite{finiteT}, is

\be 
Z=\int {\cal D}\chi_1 {\cal D}\chi_2 e^{-S_{\rm eff}}
\ee

\noindent where 
\be
S_{\rm eff}= \int\,d\tau\,L_{\rm eff}=\int\,d\tau\,[\frac{1}{2}(\dot{\chi}_1\,{\chi_1}  +\dot{\chi}_2 \,{\chi_2} )-{\rm i} \epsilon \chi_1 \chi_2]
\ee

Performing the  Fourier transformation 
\be
\chi_i(\tau)=\sum_{{\rm i}\omega_n} e^{{\rm i}\omega_n \tau} \chi_i({\rm i}\omega_n),
\ee

\noindent where ${\rm i} \omega_n$ are fermionic Matsubara frequencies, $L_{\rm eff}$ can be written
as

\be
\label{LefMat}
L_{\rm eff}= \frac{1}{2}\sum_{{\rm i}\omega_n} \sum_{i,j} \chi_i({\rm i}\omega_n) D^{-1}_{ij}({\rm i}\omega_n) \chi_j(-{\rm i}\omega_n)
\ee

\noindent where 

%\begin{widetext}
\begin{equation} 
\label{Propagm1}
D^{-1}_{ij}= 
\left(
\begin{array}{llllll}
 {\rm i}\omega_n & -{\rm i} \epsilon \\
{\rm i} \epsilon&  {\rm i}\omega_n
\end{array}
\right) \; ,
\end{equation}
%\end{widetext}

\noindent and $i$ and $j$ take values 1 and 2. 

Once the effective Lagrangian is defined, the Feynman rules 
can be read from $L_{\rm eff}$\cite{hooft73}: The inverse of the quadratic 
parts in the fields define the propagator and the remaining terms 
the interaction vertices.    

From (\ref{LefMat}) and (\ref{Propagm1}) 
the $2\times2$ fermionic propagator $D_{ij}({\rm i}\omega_n)$ is:

%\begin{widetext}
\begin{equation} 
\label{Propag}
D_{ij}({\rm i}\omega_n)= \frac{1}{({\rm i}\omega_n-\epsilon)({\rm i}\omega_n+\epsilon)}  
\left(
\begin{array}{llllll}
 {\rm i}\omega_n & {\rm i} \epsilon \\
-{\rm i} \epsilon&  {\rm i}\omega_n
\end{array}
\right)  \; ,
\end{equation}
%\end{widetext}

Note that due to the real nature of the variables $\chi$'s the propagator must be defined without  
the factor $1/2$ in (\ref{LefMat}).  

$D_{ij}({\rm i}\omega_n)$ must be associated with the propagator 
$\langle T(\hat{\chi}_i(\tau) \hat{\chi}_j(0)\rangle $ for the Majorana fermions in the ${\rm i}\omega_n$ space. 
The average $\langle \chi_1 \chi_2\rangle $ needed for the calculation of $\langle H\rangle $ is 

\be
\langle \chi_1 \chi_2\rangle =\langle T \chi_1(\tau=0^+) \chi_2(0)\rangle =\sum_{{\rm i}\omega_n} e^{{\rm i}\omega_n 0^+} D_{12}({\rm i}\omega_n)
\ee

After performing the sum over the fermionic Matsubara frequencies we obtain

\be
\label{ave}
\langle \chi_1\chi_2\rangle  =\sum_{{\rm i}\omega_n} \frac{{\rm i} \epsilon}{({\rm i}\omega_n-\epsilon)({\rm i}\omega_n+\epsilon)}=
-\frac{{\rm i}}{2}\,\tanh(\frac{\epsilon}{2k_BT}),
\ee

Thus, (\ref{ave}) leads to the correct result for $\langle H\rangle $.  

\subsection{Some characteristics of the Majorana propagator}

From (\ref{Propag}) we can also reconstruct the usual fermionic propagator  
$G(\tau)=\langle T\hat{c}(\tau) \hat{c}^\dag(0)\rangle $\cite{auerbach94}. 
From the relation between $\hat{c}$ and the $\hat{\chi}$'s we have 

\be
G(\tau)=\langle T\hat{\chi}_1(\tau) \hat{\chi}_1(0)\rangle +\langle T\hat{\chi}_2(\tau) \hat{\chi}_2(0)\rangle +
{\rm i} \langle T\hat{\chi}_1(\tau) \hat{\chi}_2(0)\rangle 
-{\rm i} \langle T\hat{\chi}_2(\tau) \hat{\chi}_1(0)\rangle 
\ee

In the Matsubara space, and using the notation of (\ref{Propag}), we get

\be
G({\rm i}\omega_n)=D_{11}({\rm i}\omega_n)+D_{22}({\rm i}\omega_n)+D_{12}({\rm i}\omega_n)+D_{21}({\rm i}\omega_n)=\frac{1}{({\rm i}\omega_n-\epsilon)}
\ee

\noindent which is the propagator or the Green function for usual fermions\cite{auerbach94}. 

Finally, we note that the propagators for $\chi_1$ and $\chi_2$, $D_{11}({\rm i}\omega_n)$  and $D_{22}({\rm i}\omega_n)$  
respectively, are odd in ${\rm i}\omega_n$ as discussed in \cite{huang15}. 
In contrast, the off-diagonal elements $D_{12}({\rm i}\omega_n)$  and $D_{21}({\rm i}\omega_n)$ are even in ${\rm i}\omega_n$.
These off-diagonal elements can be though also as coming from the coupling between 
Majorana fermions ( see also \cite{huang15} for discussions).
The two Majorana propagators $D_{11}({\rm i}\omega_n)$  and $D_{22}({\rm i}\omega_n)$ 
resonate at $\epsilon$ and $-\epsilon$. These two poles are linked to the presence of the 
off-diagonal elements of $D_{ij}$, i.e., to the coupling between the Majorana fermions.
Without the coupling between $\chi_1$ and $\chi_2$ the corresponding propagators have 
zero energy modes.

\section{Discussion and conclusion}

We have shown that models containing Majorana fermions can be identified as constrained systems.
Applying the Faddeev-Jackiw method, developed originally for constrained field theories, we have 
derived a proper effective classical Lagrangian written in terms of Grassmann variables, which  
are associated with Majorana fermions at the classical level. This effective Lagrangian leads to generalized 
Faddeev-Jackiw brackets, the equivalent to the Poison brackets of the  classical mechanics, which at the 
quantum level reproduces the noncanonical commutation rules for Majorana fermions. 
Based on this effective classical Lagrangian, the path integral was defined.   
In order to show the validity of this path integral we have applied it to an exactly solvable problem, finding
the correct result. Besides of that, this calculation is also useful to show how to perform systematic 
calculations in a straightforward way. This is important considering the present interest in this topic in the context of 
condensed matter physics, where different analytical and numerical methods are proposed.       
Finally, the method presented here may be applied to systems involved interacting Majorana fields. 
For instance, a Hubbard-Stratonovich transformation \cite{auerbach94} leads to an effective theory which is 
quadratic in the original fermions, and these fermions interact with a bosonic decoupling field. 
The quadratic part in the Majorana fields, which contains information of the original interacting problem, 
defines a propagator of the same nature of that discussed here.

\vspace*{0.5cm}
\par
\noindent
{\bf Acknowledgements}
\par
\noindent
The author thanks to M. Bejas, A. Foussats, L. Manuel, A. Muramatsu, and H. Parent for discussions.  
The author wishes to dedicate this paper to the memory
of his friend and colleague Prof. Alejandro Muramatsu,
who did not live to see several of our discussions
written in present paper


\begin{thebibliography}{10}

\bibitem{majorana37} E. Majorana, Nuovo Cimento {\bf 5}, 171 (1937).

\bibitem{aste09} A. Aste, Symmetry {\bf 2}, 1776 (2010). 

\bibitem{elliott15}  S.R. Elliott and M. Franz, Rev. Mod. Phys. {\bf 87}, 137 (2015).

\bibitem{wilczek09} F. Wilczek, Nature Physics {\bf 5}, 614 (2009). 

\bibitem{kitaev03} A.Y. Kitaev, Ann. Phys. {\bf 303}, 2 (2003). A. Y. Kitaev, 
Ann. Phys. {\bf 321}, 2 (2006).

\bibitem{karnaukhov12} I. Karnaukhov, J. Phys. A: Math, Theor. {\bf 45}, 444018 (2012).

\bibitem{nadj14} S. Nadj-Perge {\it et al.}, Science {\bf 346}, 6209 (2014).

\bibitem{nilsson13}
J. Nilsson and M. Bazzanella, Phys. Rev. B {\bf 88}, 0455112  (2013).

\bibitem{perelomov86}
A. Perelomov, {\em Generalized Coherent States and Their Applications}
  (Springer-Verlag, ADDRESS, 1986).

\bibitem{shankar11} R. Shankar
and A. Vishwanath, Phys. Rev. Lett. {\bf 107}, 106803 (2011).
    
\bibitem{Dirac64}
P.~A.~M. Dirac, {\em Lecture on Quantum Mechanics} (Yeshiva Univ. Press,
  ADDRESS, 1964).
  
\bibitem{Faddeev88}
L. Faddeev and R. Jackiw, Phys. Rev. Lett. {\bf 60},  1692  (1988).

\bibitem{Govaerts90}
J. Govaerts, Int. J. Mod. Phys. A {\bf 5},  3625  (1990).

\bibitem{hooft73} G. 't Hooft and M. Velman, Diagramar (CERN, Geneve, 1973).

\bibitem{foussats99a}
A. Foussats, A. Greco, and O. Zandron, Int. J. Theor. Phys. {\bf 38},  1439
  (1999).

\bibitem{foussats99b}
A. Foussats, A. Greco, and O. Zandron, Ann. Phys., NY {\bf 275},  238  (1999).

\bibitem{foussats11}
A. Foussats, A. Greco, and A. Muramatsu, Nuclear Phys. B {\bf 842}, 225 (2011).

\bibitem{Barcelo92}
J. Barcelos-Neto and C. Wotzasek, Int. J. Mod. Phys. A {\bf 7},  4981  (1992).

\bibitem{berezin} A. Berezin, {\em The Method of Second Quantization}, Academic Press, (1966)..

\bibitem{Gitman90}
D. Gitman and I. Tyutin, {\em Quantization of Fields with Constraints}
  (Springer, Berlin, ADDRESS, 1990).

\bibitem{senjanovic} P. Senjanovic, Ann. Phys. NY {\bf 100}, 227 (1976).

\bibitem{auerbach94}
A. Auerbach, {\em Interacting Electrons and Quantum Magnetism}
  (Springer-Verlag, ADDRESS, 1994).
  
\bibitem{finiteT} C. Bernard, Phys. Rev. D {\bf 9}, 3312 (1974). 

\bibitem{huang15} Z. Huang, P. Wolfle, and A.V. Balatsky, Phys. Rev. B. {\bf 92}, 121404(R) (2015)

\end{thebibliography}
\end{document}